\documentclass[oldversion]{aa}	
\usepackage{txfonts}
\usepackage{graphicx}
\def\Msolar{\hbox{${\rm M}_\odot$}}

\begin{document} 

\title{NGC 2298: a globular cluster on its way to disruption\thanks{Based
on observations with the NASA/ESA {\it Hubble Space Telescope},
obtained at the Space Telescope Science Institute, which is operated by
AURA, Inc., under NASA contract NAS5-26555}}

\author{Guido De Marchi \inst{1} and Luigi Pulone \inst{2} }

\offprints{Guido De Marchi}

\institute{ESA, Space Science Department, Keplerlaan
1, 2200 AG Noordwijk, Netherlands, gdemarchi@rssd.esa.int 
\and
INAF, Osservatorio Astronomico di Roma, Via di Frascati 33, 00040 Monte
Porzio Catone, Italy, pulone@mporzio.astro.it}

\date{Received 9.11.2006; Accepted 29.11.2006}
\titlerunning{Tidal disruption of NGC\,2298}
\authorrunning{De Marchi et al.}

\abstract{ 

We have studied the stellar main sequence (MS) of the globular cluster
NGC\,2298 using deep HST/ACS observations in the F606W and F814W bands
covering an area of $3\farcm4 \times 3\farcm4$ around the cluster
centre or about twice the cluster's half-mass radius. The
colour-magnitude diagram that we derive in this way reveals a
narrow and well defined MS extending down to the $10\,\sigma$ detection
limit at $m_{\rm 606} \simeq 26.5$, $m_{\rm 814} \simeq 25$,
corresponding to  stars of $\sim 0.2$\,\Msolar. The luminosity function
(LF) obtained with these data, once corrected for the limited effects
of photometric incompleteness, reveals a remarkable deficiency of
low-mass stars as well as a radial gradient, in that the LF becomes
progressively steeper with radius. Using the mass--luminosity relation
appropriate for the metallicity of NGC\,2298, we derive the cluster's
global mass function (GMF) by using a multi-mass Michie--King model.
Over the range $0.8 - 0.2$\,\Msolar, the number of stars per unit mass
decreases following a power-law distribution of the type $dN/dm \propto
m^{0.5}$, where, for comparison, typical halo clusters have $dN/dm \propto
m^{-1.5}$. If the IMF of NGC\,2298 was similar to that of other metal
poor halo clusters, like e.g. NGC\,6397, the present GMF that we
obtain implies that this object must have lost of order 85\,\% of its
original mass, at a rate much higher than that suggested by current
models based on the available cluster orbit. The latter may, therefore,
need revision. 

}

\maketitle

\section{Introduction}

Observations over the past years have revealed a growing number of
globular clusters (GCs) severely depleted of low-mass stars. These
include NGC\,6712 (De Marchi et al. 1999; Andreuzzi et al. 2001),
Pal\,5 (Koch et al. 2004), NGC\,6218 (De Marchi et al. 2006) and
NGC\,6838 (Pulone et al. 2007). Their stellar mass function (MF) shows
that the number of stars per unit mass does not increase towards a peak
near $0.3$\,\Msolar, as is typical of GCs (see Paresce \& De Marchi
2000, hereafter PDM00), but stays roughly constant and in most cases
decreases with mass. These findings are interpreted as being the result
of mass loss through the evaporation of stars, most likely induced or
enhanced by the tidal field of the Galaxy. 

It has long been predicted that GCs in a tidal environment should
suffer a more rapid and severe mass loss than isolated systems (H\'enon
1961, 1965) and in the past decade and a half a sizeable amount of work
has been made to quantify the strength and extent of these effects, by
means of increasingly realistic dynamical simulations. The most
relevant papers in this field are those of Aguilar, Hut \& Ostriker
(1989), Gnedin \& Ostriker (1997), Vesperini \& Heggie (1997), Dinescu
et al. (1999b) and Baumgardt \& Makino (2003). While the earlier works
mainly aim to analyse the GC system as a whole, to try and understand
which fraction of the original population of GCs we see today (Aguilar
et al. 1989; Gnedin \& Ostriker 1997), the most recent studies
incorporate up to date orbital parameters and assign specific
probability of disruption or remaining lifetimes to individual objects
(Dinescu et al. 1999b; Baumgardt \& Makino 2003). 

There is general consensus that the combined effect of the internal
dynamical evolution, via two-body relaxation, and the interaction with
the Galaxy and ensuing tidal stripping is such that GCs will
preferentially lose low-mass stars in the course of their life.
Therefore, the distribution of stellar masses will progressively depart
from that of the initial mass function (IMF) and the ratio of lower and
higher mass stars will tend to decrease over time. In other words, the
global mass function (GMF), i.e. the MF of the cluster as a whole, will
become flatter, particularly at the low mass end (Vesperini \& Heggie
1997). Consequently, GCs with a higher probability of disruption (no
matter whether dominated by internal or external processes) should
display a more depleted, i.e. flatter GMF. 

At least qualitatively, this picture is consistent with the GMF of the 
four severely depleted clusters mentioned above, in that their
predicted time to disruption ($T_{\rm dis}$) is typically shorter than
that of the average cluster (Gnedin \& Ostriker 1997; Dinescu et al.
1999b; Baumgardt \& Makino 2003). Unfortunately, when one attempts to
establish a quantitative relationship or correlation between the
observed GMF shape and the predicted value of $T_{\rm dis}$ of these
and any other GCs, one is faced with a serious mismatch, in that some
clusters with short $T_{\rm dis}$ have steep GMFs and vice-versa
(PDM00; De~Marchi et al. 2006). Furthermore, the value of $T_{\rm dis}$
predicted by different authors for the same cluster can vary
considerably, owing also to different or incorrect assumptions on
the form of the Galactic potential or on the details of the cluster's
orbit. This suggests that the evolution of the GC system and the
interplay between the Galactic tidal field and the internal dynamical
evolution of individual clusters are not yet fully understood.

However, an interesting pattern is observed in that all depleted
clusters so far known have relatively low central concentration (De
Marchi et al. 2006), as  defined by the King central concentration
parameter $c=\log \, (r_{\rm t}/r_{\rm c})$, namely the logarithmic
ratio of the tidal radius $r_{\rm t}$ and core radius $r_{\rm c}$. This
finding is puzzling because the selective removal of low-mass
stars requires the combined effect of tidal stripping, which removes
stars from the outskirts of the cluster, and of two-body relaxation,
which  causes the loss of stars via evaporation but also, more
importantly, feeds low-mass stars to the cluster's periphery. The
apparent anomaly is that the two-body relaxation process should also
drive the cluster to higher values of $c$ and eventually core collapse
(Spitzer 1987), contrary to what is observed. On the other hand, the
current sample of four clusters is too small to assess how significant
is the observed trend and additional studies of low- and
intermediate-concentration clusters are needed to understand its
relevance for the dynamical evolution of GCs. 

Therefore, in this paper we undertake a detailed study of the MF of
NGC\,2298, a cluster of intermediate concentration ($c=1.3$; Harris
1996) recently observed with the HST. This is a particularly
interesting object since its low metallicity ([Fe/H]$=-1.85$; Harris
1996) and orbital parameters (Dinescu et al. 1999a) are typical of a
halo cluster that should have experienced a rather mild interaction
with the Galactic tidal field. Its GMF should, therefore, show little
signs of low-mass star depletion. On the other hand, if the observed
trend between central concentration and GMF shape is real, then
NGC\,2298 should have a relatively shallow GMF slope, in light of its
intermediate $c$ value.  

The structure of the paper is as follows. Section\,2 describes briefly
NGC\,2298 and the new HST data, along with the reduction process; the
photometry is presented in Section\,3, while Sections\,4 and 5 are
devoted to the luminosity function (LF) and MF variations throughout
the cluster, respectively; a discussion and the conclusions follow in
Section\,6.

\begin{figure*}
\centering
\resizebox{\hsize}{!}{\includegraphics{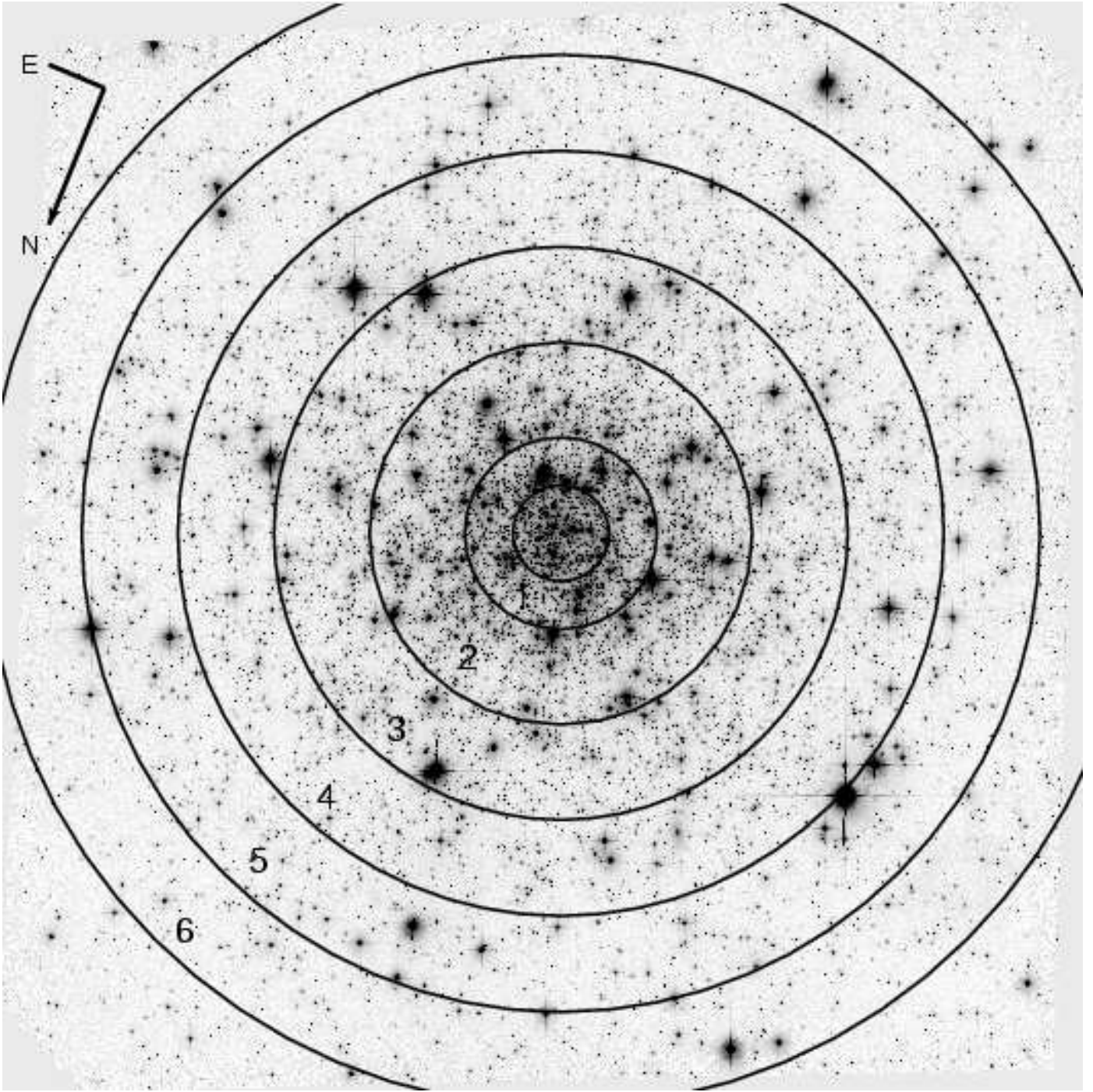}}
\caption{Negative image of the cluster NGC\,2298 through the F606W
band.  The field spans $3\farcm4$ on a side. The circles marked on the
frame define the six annuli in which we have divided our photometry,
as explained in Section\,3.}
\label{fig1}
\end{figure*}

\section{Observations and data analysis}

NGC\,2298 was observed with the Advanced Camera for Surveys (ACS; see
Ford et al. 2003; Pavlovsky et al. 2006) on board the HST on 2006, June
12 as part of the treasury programme entitled ``An ACS Survey of
Galactic Globular Clusters'' (P.I. Ata Sarajedini). A series of five
dithered exposures lasting 350\,s each (plus an additional short
exposure of 20\,s duration) were collected through the F606W and F814W
filters, for a total exposure time of 1770\,s in each band. The
telescope was pointed at the nominal centre of NGC\,2298 at
RA(J2000)$=06^{\rm h} 48^{\rm m} 59^{\rm s}5$ and DEC(J2000)$=-36^{\rm
o} 00\arcmin 30\arcsec$, corresponding to Galactic coordinates
$l=245.6$ and $b=-16.0$. The field of view of the ACS Wide Field
Channel, used for these observations, spans about $3\farcm4$ on a side
and reaches out to more than twice the cluster's half-mass radius
($r_{\rm h}=0\farcm78$; Harris 1996). 

The raw data were retrieved from the HST archive after having been
processed through the standard ACS pipeline calibration, which also took
care of correcting the geometric distortion and calibrating the
photometric zero point (Sirianni et al. 2005). The procedure also
registered and combined the images in the same band. In Figure\,1 we
show a negative image of the final combined F606W frame. 

Star detection and identification was done by running the automated
IRAF {\em daofind} routine on the F606W band combined frame, with the
detection threshold set at $10\,\sigma$ above the local background
level. This conservative choice for the detection threshold stems from
the fact that in this work we are not specifically interested in
detecting the faintest possible stars, but rather in mapping as
uniformly as possible low-mass MS stars throughout the cluster. While 
it would have been possible to push the detection threshold down to
$\sim 5\,\sigma$, this would have also implied a higher degree of
photometric incompleteness and, therefore, a less robust LF. On the
other  hand, owing to the large number of saturated stars in the
frames, even with our $10\,\sigma$ detection threshold a careful
inspection of all objects found by the automated routine is necessary,
since point spread function (PSF) tendrils and artefacts can easily be
misclassified as bona-fide stars. Fully or partly resolved objects,
namely background galaxies, were also eliminated from the catalogue
during this procedure. The final star list obtained in this way
includes $11,350$ unresolved objects, comprising both cluster and field
stars. 

Except for the innermost $\sim 10\arcsec$ radius, which we did not
study, crowding is not severe and stellar magnitudes can be accurately
estimated with aperture photometry. We used the standard IRAF {\em
DAOPhot} package (Stetson et al. 1987) and, following the core
aperture photometry  technique (De Marchi et al. 1993), we set the
aperture radius to $0\farcs1$, while measuring the local background in
an annulus extending from $0\farcs15$ to $0\farcs25$ around each object. 

Since the PSF of the WFC varies considerably across the field of view
(Sirianni et al. 2005), photometry based on PSF-fitting requires an
accurate knowledge of the PSF. Aperture photometry, on the other hand,
is less sensitive to the actual change of the PSF, provided that the
aperture is large enough to accommodate the PSF changes (Sirianni et
al. 2005). We have experimented with several choices for the aperture
and background annulus and have selected the combination indicated
above since it provides the least photometric scatter around the main
sequence (MS) in the colour--magnitude diagram (CMD; on average
$0.04$\,mag).  

Instrumental magnitudes were finally calibrated using the photometric
zero points for the ACS VEGAMAG magnitude system (Sirianni et al.
2005). Our $10\,\sigma$ detection limits correspond to magnitudes
$m_{\rm 606} \simeq 26.5$ and $m_{\rm 814} \simeq 25$, where the
relative photometric error is $\la 0.05$ mag. The absolute photometric
uncertainty is of the same order of magnitude, mainly due to the use of
a small aperture, which implies a relatively large aperture correction.
More sophisticated photometric techniques (e.g. Anderson \& King 2004),
which take into account the local variation of the PSF, might further
reduce the overall uncertainty and the scatter along the MS in the CMD.
However, the level of accuracy that we have reached is more than
adequate for the purpose of understanding the broad properties of the
stellar MF of NGC\,2298. For this reason, we decided to also ignore the
effects of the imperfect charge-transfer efficiency affecting ACS data
(Sirianni et al. 2005), which contribute to broadening the MS in the
CMD. None of the conclusions that we draw in this work are sensitive to
this effect.   

\begin{figure}
\centering
\resizebox{\hsize}{!}{\includegraphics[width=16cm]{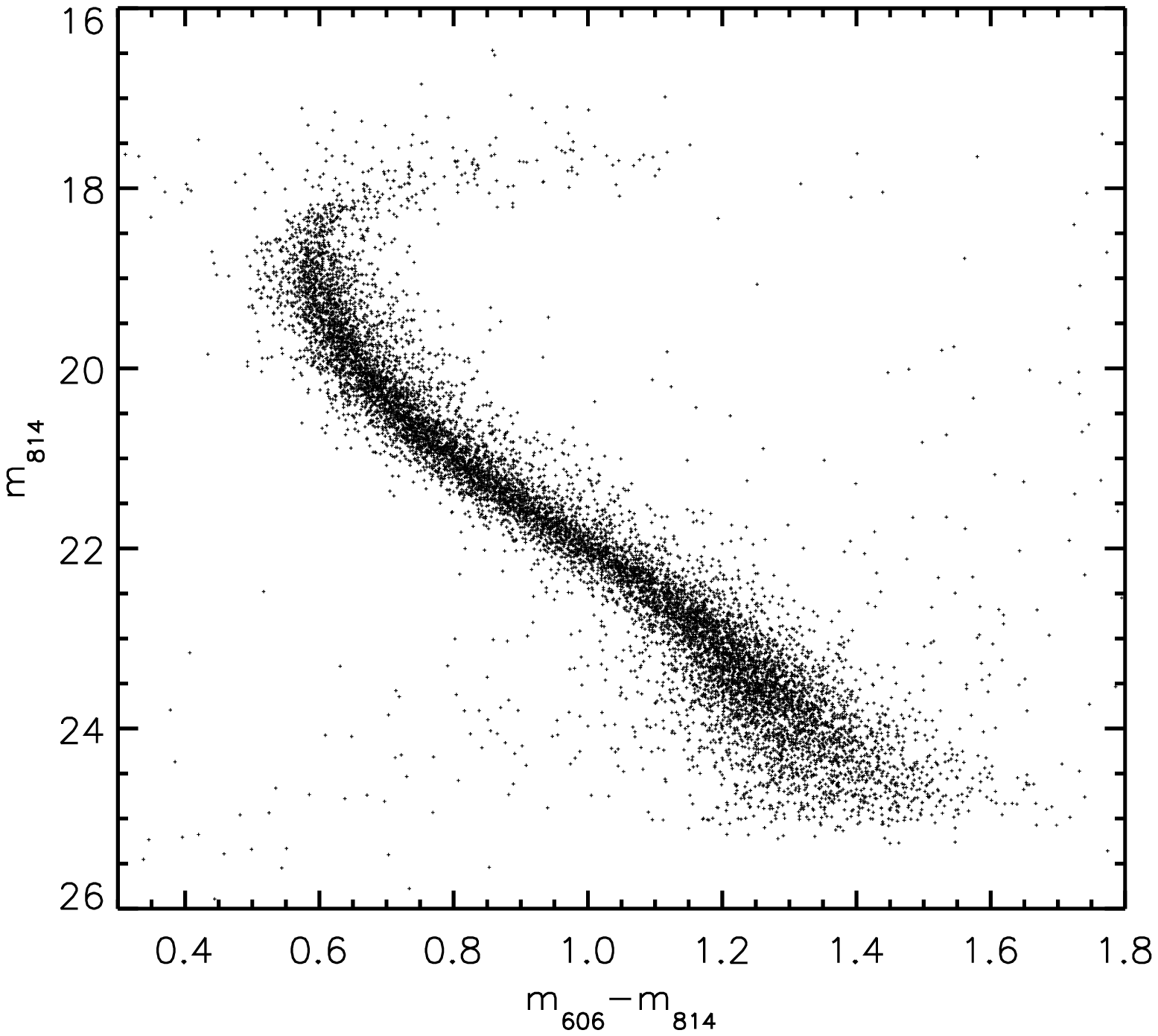}}
\caption{Colour--magnitude diagram of an area of $3\farcm4 \times
3\farcm4$ around the centre of NGC\,2298.} 
\label{fig2}
\end{figure}

\section{The colour--magnitude diagram}

The CMD of the complete set of $11,350$ objects that we have detected
is shown in Figure\,\ref{fig2}, where the cluster's MS is clearly
visible. At magnitudes brighter than $m_{\rm 814} \simeq 18.5$ the CMD
reveal stars evolving off the MS (sub-giant branch), but their
photometry is increasingly unreliable, due to saturation. Here we
concentrate on the cluster's MS, which is narrow and well defined from
the turn-off at $m_{\rm 814}=18.8$, where the photometric error is
small ($\sigma_{\rm 606} \simeq \sigma_{\rm 814} \simeq 0.01$), through
to $m_{\rm 814} \simeq 25$, where the error on the magnitude grows to
$\sigma_{\rm 606} \simeq \sigma_{\rm 814} \simeq 0.04$. 

Owing to the relatively low Galactic latitude of NGC\,2298
($b=-16^\circ$), some field star contamination has to be expected in
the CMD of Figure\,\ref{fig2}. This can be seen at $m_{\rm 814} \ga
22$, where an increasing number of objects appears on both sides of the
MS, with  colours too red or too blue for them to be bona-fide MS
stars, given the typical magnitude uncertainty at this brightness
($\sigma_{\rm 606} \simeq \sigma_{\rm 814} \simeq 0.04$).  

As a first attempt to roughly estimate the relevance of this
contamination, we turned to the Galaxy model of Ratnatunga \& Bahcall
(1985), which predicts about 17 field stars per arcmin square towards
the direction of NGC\,2298 down to magnitude $V=27$, with about half of
them in the range $24 < V < 27$. This would correspond to $\sim 200$
contaminating field stars in our field of view down to $V=27$, of which
$\sim 100$ brighter than $V=24$. The CMD of Figure\,\ref{fig2} has of
order 3,950 stars with $m_{\rm 606} \ge 24$, before any correction for
incompleteness, and there are approximately $7,400$ stars brighter than
that magnitude. Therefore, it would seem perfectly unnecessary to take
field star contamination into account.

On the other hand, since field star contamination affects more
prominently the lower end of the MS, which is most relevant to our
investigation, we decided to remove it at all magnitudes, following a
statistical approach. To this aim, we made use of the colour
information in the CMD and applied the $\sigma$-clipping criterion
described by De Marchi \& Paresce (1995) to identify the possible
outliers. In practice, from the CMDs of Figure\,\ref{fig2} we counted
the objects in each $0.5$ mag bin and within $\pm 2.5$ times the colour
standard deviation around the MS ridge line and rejected the rest as
field objects. This procedure is iterative and at each step the colour
of the ridge line and its associated standard deviation are
recalculated. Convergence is reached, usually within a few iterations,
when all stars whose colour differs from the average by more than $\pm
2.5$ times the MS standard deviation have been removed. In this way we
identified about $320$ field stars, of which about $150$ fainter than
$m_{\rm 606} \simeq 24$ (or $m_{\rm 814} \simeq 22$). These values are
within a factor of two of those predicted by the model of Ratnatunga \&
Bahcall (1985) and as such are in surprisingly good agreement, given
the large uncertainty affecting the Galaxy model of Bahcall \& Soneira
(1984) for $10^\circ < |b| < 20^\circ$, due to the lack of direct
observational calibration. 

The number of putative field stars, determined with our procedure, is
shown in Table\,\ref{tab1} as a function of the magnitude, together
with the average MS colour (ridge line) and associated standard
deviation. We stress here that field stars account for an insignificant
fraction ($\sim 3\,\%$) of the bona-fide MS stars and that, in each
magnitude bin, their number is smaller than the statistical uncertainty
associated with the counting process. In the following, we will
consider only bona fide MS stars as defined in this way but, as will
become clear in the rest of the paper, our determination of the MF of
NGC\,2298 would remain unchanged if we had decided to ignore field star
contamination and treat all stars as cluster members.

\begin{table} 
\centering
\caption{Average main sequence fiducial points, colour width and number
of field stars.}
\begin{tabular}{cccc}
\hline 
 $m_{\rm 814}$ & $m_{\rm 606}-m_{\rm 814}$ & $\sigma_{\rm 606-814}$ &
$N_{\rm F}$ \\ 
\hline
 18.75 &  0.592 &  0.035 & 12 \\
 19.25 &  0.608 &  0.035 & 20 \\
 19.75 &  0.642 &  0.035 & 28 \\
 20.25 &  0.692 &  0.037 & 24 \\
 20.75 &  0.765 &  0.041 & 22 \\
 21.25 &  0.850 &  0.043 & 27 \\
 21.75 &  0.952 &  0.050 & 29 \\
 22.25 &  1.057 &  0.050 & 27 \\
 22.75 &  1.152 &  0.050 & 23 \\
 23.25 &  1.223 &  0.050 & 29 \\
 23.75 &  1.274 &  0.061 & 36 \\
 24.25 &  1.332 &  0.079 & 25 \\
 24.75 &  1.385 &  0.097 & 15 \\ 
\hline
\end{tabular}
\vspace{0.5cm}
\label{tab1}
\end{table}

\begin{figure}
\centering
\resizebox{\hsize}{!}{\includegraphics{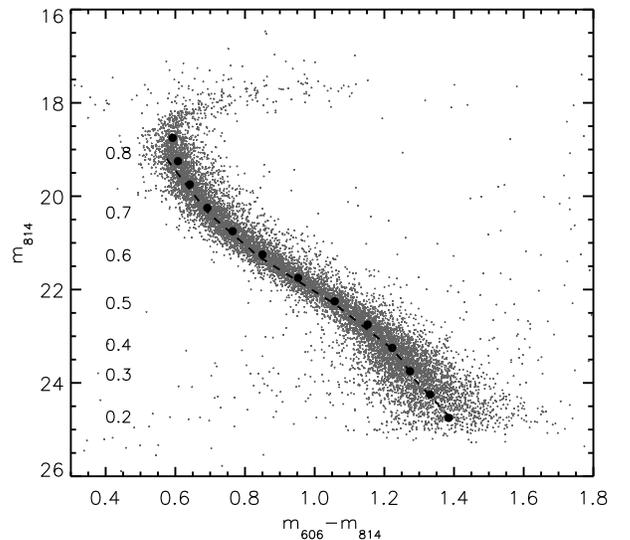}}
\caption{Main sequence ridge line of NGC\,2298 (filled circles)
compared with the theoretical isochrones of Baraffe et al. (1997;
dashed line) for an age of 10\,Gyr and metallicity $[M/H]=-1.5$, in the
magnitude system of the ACS (see text).}
\label{fig3}
\end{figure}

Figure\,\ref{fig3} shows, traced over the same CMD of
Figure\,\ref{fig2}, the MS ridge line (filled circles) obtained by
applying the sigma-clipping method explained above. The dashed line in
the same figure corresponds to the theoretical isochrone computed by
Baraffe et al. (1997) for a 10\,Gyr old cluster with metallicity
$[M/H]=-1.5$ and Helium content $Y=0.25$, as is appropriate for
NGC\,2298 (McWilliam, Geisler \& Rich 1992; Testa et al. 2001). We have
assumed a distance modulus $m-M=15.15$ and colour excess $E(B-V)=0.14$
from Harris (1996). The models of Baraffe et al. (1997) are calculated
for the F606W and F814W filters of the WFPC2 camera, which do not
exactly match those on board the ACS. The magnitude difference between
the two cameras in the same band is a function of the effective
temperature of the star (hence of its mass) and can be approximately
estimated by means of the HST synthetic photometric package ``Synphot''
(Laidler et al. 2005). To calculate the correction factors, we have
used synthetic atmosphere models from the ATLAS9 library of Kurucz
(1993) with effective temperature and surface gravity matching those of
the models of Baraffe et al. (1997). The resulting differences, in the
F606W band, range from $0.028$\,mag for stars of $0.8$\,\Msolar to
$0.074$\,mag for stars of $0.2$\,\Msolar. In the F814W band, the
differences are smaller and range from $0.007$\,mag to $0.036$\,mag
over the same mass range. The dashed line in Figure\,\ref{fig3} shows
the effect of translating the Baraffe et al. (1997) models to the ACS
magnitude system, over the mass range $0.2 - 0.8$\,\Msolar (the mass
points are indicated on the left-hand side). The agreement with the MS
ridge line is very good and convinced us that we can use the translated
models to convert magnitudes into masses.

\section{The luminosity and mass function} 

The LF of MS stars is derived from the CMD of Figure\,\ref{fig2} as
part of the same sigma-clipping process that rejects field stars, since
all objects that are not rejected are by definition bona-fide MS
stars. Given the richness of the photometry and the wide field covered
by the data, reaching out to more than twice the cluster's half-mass
radius ($r_{\rm h}=0\farcm78$; Harris 1996), it is possible to study
with accuracy the variation of the LF with radius. This is an essential
step in securing a solid GMF.

To this aim, we have grouped our photometric catalogue to form a series
of 6 concentric annuli, centered on the nominal cluster centre and
expending out to $120\arcsec$, with a step of $20\arcsec$. Since the
stellar density increases considerably towards the cluster's centre, we
did not consider the innermost $10\arcsec$ radius where the photometric
completeness is very low due to the many bright stars. For simplicity,
hereafter we refer to these regions as ``rings'' 1 through to 6, as
marked in Figure\,\ref{fig1}. They contain, respectively, 1653, 2609,
3016, 2827, 1946, and 566 objects.\footnote{We note here that, due to
the $\sim 20\arcsec$ gap between the two detectors of the ACS/WFC and
in light of the dithering pattern used for these observations, the
cluster centre and the regions to the right and to the left of it in
the images do not have the same photometric depth as the other parts of
the frames. We have deliberately avoided these regions and, therefore,
no photometry exists for certain sections of the rings. Therefore, the
number of objects detected in each ring does not necessarily correspond
to the radial density profile.} Ring\,3 is of particular importance,
since it covers the region around the cluster's half-mass radius, where
the properties of the local MF are expected to be as close as possible
to those of the global MF (Richer et al. 1991; De Marchi \& Paresce
1995; De Marchi et al. 2000).

The large density gradient in the images implies that the completeness
of the photometry is not uniform across the field. Incompleteness is
due to crowding and to saturated stars, whose bright halo can mask
possible faint objects in their vicinity, both more likely to affect
substantially the central regions (see Figure\,\ref{fig1}). If ignored
or corrected for in a uniform way across the whole image, by applying
the same correction to all regions, these effects would bias the final
results and mimic the presence of a high degree of mass segregation.
For this reason, we conducted a series of artificial star experiments
over the regions corresponding to each  of the five individual annuli.

The artificial star tests were run on the combined images, in the F606W
band (the same used for star detection). For each $0.5$ magnitude bin
we carried out 10 trials by adding a fraction of 10\,\% of the total
number of objects (see Section 2). The artificial images were then
reduced with the same parameters used in the reduction of the
scientific images so that we could assess the fraction of objects
recovered by the procedure. The resulting photometric completeness is
given in Table\,2 for each region as a function of the $m_{\rm 814}$
magnitude.

\begin{table} 
\centering
\caption{Photometric completeness $f$ in the six individual rings.} 
\begin{tabular}{ccccccc}
\hline 
$m_{\rm 814}$ & Ring\,1 & Ring\,2 & Ring\,3 & Ring\,4 & Ring\,5 & Ring\,6 \\ 
\hline
 18.75 &   0.95 &   0.97 &   0.97 &   0.99 &   0.99 &   0.99 \\
 19.25 &   0.93 &   0.96 &   0.97 &   0.98 &   0.99 &   0.99 \\
 19.75 &   0.91 &   0.95 &   0.96 &   0.98 &   0.99 &   0.99 \\
 20.25 &   0.89 &   0.95 &   0.96 &   0.98 &   0.99 &   0.99 \\
 20.75 &   0.86 &   0.95 &   0.96 &   0.98 &   0.99 &   0.99 \\
 21.25 &   0.84 &   0.94 &   0.95 &   0.97 &   0.99 &   0.99 \\
 21.75 &   0.82 &   0.94 &   0.95 &   0.97 &   0.99 &   0.99 \\
 22.25 &   0.79 &   0.93 &   0.95 &   0.97 &   0.98 &   0.98 \\
 22.75 &   0.75 &   0.93 &   0.94 &   0.96 &   0.98 &   0.98 \\
 23.25 &   0.71 &   0.92 &   0.94 &   0.96 &   0.97 &   0.97 \\
 23.75 &   0.62 &   0.89 &   0.91 &   0.94 &   0.97 &   0.97 \\
 24.25 &   0.53 &   0.86 &   0.89 &   0.93 &   0.96 &   0.96 \\
 24.75 &   0.50 &   0.85 &   0.88 &   0.92 &   0.96 &   0.96 \\
\hline
\end{tabular}
\vspace{0.5cm}
\label{tab2}
\end{table}

\begin{figure}  
\centering
\resizebox{\hsize}{11.65cm}{\includegraphics{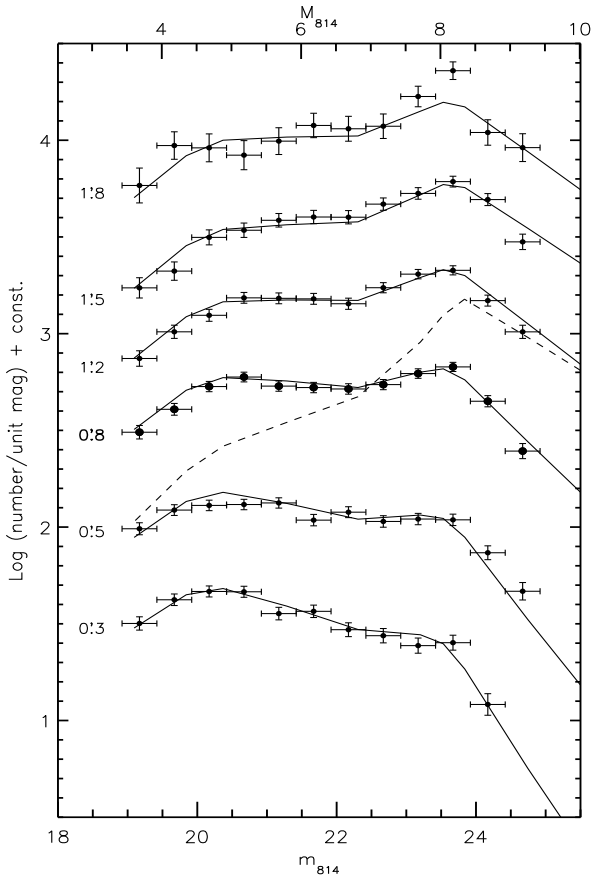}}
\caption{The circles are the LFs of the six rings, after correction for
photometric incompleteness (see Table\,2). The actual measurements from
Table\,\ref{tab3} are shifted vertically by an arbitrary amount to
increase readability. The geometric average radius of each ring is
indicated in arcmin to the left of each curve. The theoretical LFs that
best fit the data are shown as solid lines. The index of their
corresponding power-law MF is, from bottom to top, $\alpha=1.6, 1.1,
0.5, 0.1, 0$ and $-0.1$. A positive index means that the number of
stars decreases with mass. All but the innermost LF reach down to
$0.2$\,\Msolar. The dashed  line is the LF measured near the half-mass
radius of NGC\,6397 by De  Marchi et al. (2000), with arbitrary
vertical normalisation.} 
\label{fig4} 
\end{figure}

\begin{table*} 
\centering
\caption{Luminosity functions measured in each of the six rings. For 
each region, the table gives, as a function of the $m_{\rm 814}$
magnitude, the number of stars per half-magnitude bin before ($N_{\rm
o}$) and after ($N$) completeness correction and the uncertainty 
($\sigma_{\rm N}$) on $N$.}

\begin{tabular}{c@{\hspace{0.7cm}}
               ccc@{\hspace{0.7cm}}ccc@{\hspace{0.7cm}}ccc@{\hspace{0.7cm}}
               ccc@{\hspace{0.7cm}}ccc@{\hspace{0.7cm}}ccc}
\hline 
  \multicolumn{1}{c}{\,} 
     &  \multicolumn{3}{c}{Ring\,1} &  \multicolumn{3}{c}{Ring\,2}  
     &  \multicolumn{3}{c}{Ring\,3} &  \multicolumn{3}{c}{Ring\,4} 
     &  \multicolumn{3}{c}{Ring\,5} &  \multicolumn{3}{c}{Ring\,6}\\ 
 $m_{\rm 814}$ &  $N_{\rm o}$ & $N$ & $\sigma_{\rm N}$  
     &    $N_{\rm o}$ & $N$ & $\sigma_{\rm N}$ 
     &    $N_{\rm o}$ & $N$ & $\sigma_{\rm N}$ 
     &    $N_{\rm o}$ & $N$ & $\sigma_{\rm N}$
     &    $N_{\rm o}$ & $N$ & $\sigma_{\rm N}$ 
     &    $N_{\rm o}$ & $N$ & $\sigma_{\rm N}$ \\
\hline
 18.75 &  78 &   82 &    9 &  96 &   99 &   10 & 101 &  103 &   10 &  84 &   85 &    9 &  60 &   60 &    7 &  11 &   11 &    3 \\
 19.25 & 154 &  166 &   13 & 188 &  195 &   15 & 146 &  151 &   13 & 121 &  123 &   11 &  64 &   64 &    8 &  25 &   25 &    5 \\
 19.75 & 190 &  209 &   16 & 228 &  239 &   17 & 196 &  203 &   15 & 154 &  157 &   13 &  86 &   86 &    9 &  35 &   35 &    6 \\
 20.25 & 197 &  222 &   16 & 229 &  241 &   17 & 249 &  259 &   18 & 195 &  199 &   15 & 124 &  125 &   11 &  31 &   31 &    5 \\
 20.75 & 191 &  221 &   16 & 244 &  257 &   18 & 285 &  298 &   19 & 237 &  242 &   17 & 141 &  142 &   12 &  42 &   42 &    6 \\
 21.25 & 156 &  185 &   15 & 246 &  260 &   18 & 259 &  271 &   18 & 224 &  230 &   16 & 135 &  136 &   12 &  37 &   37 &    6 \\
 21.75 & 139 &  169 &   14 & 199 &  211 &   16 & 244 &  256 &   17 & 223 &  229 &   16 & 160 &  161 &   13 &  40 &   40 &    6 \\
 22.25 & 117 &  148 &   13 & 215 &  230 &   16 & 228 &  240 &   17 & 224 &  231 &   16 & 155 &  157 &   13 &  43 &   43 &    6 \\
 22.75 & 100 &  133 &   12 & 193 &  208 &   15 & 261 &  276 &   18 & 255 &  264 &   18 & 186 &  190 &   14 &  45 &   46 &    6 \\
 23.25 &  79 &  111 &   11 & 198 &  216 &   16 & 294 &  314 &   20 & 324 &  338 &   21 & 194 &  200 &   15 &  73 &   75 &    8 \\
 23.75 &  80 &  129 &   12 & 183 &  206 &   15 & 287 &  314 &   20 & 293 &  311 &   20 & 229 &  237 &   17 &  73 &   75 &    9 \\
 24.25 &  24 &   45 &    7 & 113 &  131 &   12 & 178 &  200 &   15 & 208 &  224 &   16 & 171 &  177 &   14 &  41 &   42 &    6 \\
 24.75 & --- &  --- &  --- &  70 &   82 &    9 &  98 &  111 &   11 & 124 &  134 &   12 &  99 &  103 &   10 &  27 &   28 &    5 \\
\hline
\end{tabular}
\vspace{0.5cm}
\label{tab3}
\end{table*}

The LF of MS stars in each ring is shown graphically in
Figure\,\ref{fig4}, in units of apparent and absolute magnitude in the
F814W band. Table\,\ref{tab3} gives the five LFs, before and after
correction for photometric incompleteness, and the corresponding rms
errors coming from the Poisson statistics of the counting process (only
for the LF corrected for incompleteness). All values have been rounded
off to the nearest integer. The LFs are remarkably flat, but a
progressive steepening is evident with increasing radius. This trend is
not the result of decreasing photometric completeness since only
data-points with an original associated photometric completeness in
excess of 85\,\% (50\,\% for Ring\,1) are shown in Figure\,\ref{fig4},
as witnessed by Table\,\ref{tab2}. 

It is instructive to compare the LF of Ring\,3 (at the cluster's
half-mass radius) with that of NGC\,6397, shown as a dashed line in
Figure\,\ref{fig4}. Both are halo clusters with low metallicity
([Fe/H]$=-1.95$ for NGC\,6397 and [Fe/H]$=-1.85$ for  NGC\,2298; Harris
1996), but while near the half-mass radius of NGC\,6397 the number of
stars per unit magnitude grows by a factor of about four from the
turn-off luminosity to the LF peak at $M_{\rm 814} \simeq 8$  (King et
al. 1998), for NGC\,2298 it only changes by a factor of two. NGC\,2298
is therefore seriously lacking low-mass stars with respect to
NGC\,6397.

Since the MS of NGC\,2298 is well fitted by the theoretical isochrones
of  Baraffe et al. (1997), after conversion to the ACS photometric
system (see Figure\,\ref{fig3}), we can use the associated
mass--luminosity (M--L) relationships to convert magnitudes to masses.
In order to keep observational errors clearly separated from
theoretical uncertainties, we prefer to fold a model MF through the
M--L relationship and compare the resulting model LF with the
observations. The solid lines in Figure\,\ref{fig4} are the theoretical
LFs obtained by multiplying a simple power-law MF of the type $dM/dm
\propto m^\alpha$ by the derivative of the M--L relation. Over the mass
range ($0.2 - 0.8$\,\Msolar) spanned by these observations a simple
power-law is a valid representation of the MF, although at lower masses
the MF progressively departs from a power-law and this simplifying
assumption is no longer a good representation of the mass distribution
in GCs (PDM00; De Marchi, Paresce \& Portegies Zwart 2005).

With the adopted M--L relation, a distance modulus $m-M=15.15$ and 
colour excess $E(B-V)=0.14$ (see Section\,3), the best fitting local
power-law slopes are $\alpha=1.6, 1.1, 0.5, 0.1, 0, -0.1$ respectively
at $r= 0\farcm3, 0\farcm5, 0\farcm8, 1\farcm2, 1\farcm5, 1\farcm8$. The
fact the value of the index is in almost all cases positive implies
that the number of stars is decreasing with mass. With the notation
used here, the canonical Salpeter IMF would have $\alpha=-2.3$. For
comparison, over the same mass range spanned by the MF of Ring\,3, the
power-law index that best fits the MF of NGC\,6397 is $\alpha \simeq
-1.6$ (see dashed line in Figure\,\ref{fig4}) and that of the other 11
halo GCs in the sample studied by PDM00 is of the same order. It is,
therefore, clear that NGC\,2298 is surprisingly devoid of low-mass
stars, at least out to over twice its half-mass radius.

\section{The global mass function}

Given the large spread of MF indices that we find, ranging over a dex,
it is fair to wonder which of these MFs, if any, is representative of
the mass distribution in the cluster as a whole. It is in principle
possible (although not very likely, since the observations sample well
more than half of the cluster's population by mass), that a large
number of low-mass stars is present in the cluster, but located beyond
the radius covered by the these data. To clarify this issue, it is
useful to study the dynamical state of NGC\,2298 and investigate
whether the observed MF variation with radius is consistent with the
degree of mass segregation expected from energy equipartition due to
two-body relaxation.   

To this aim, we followed the approach outlined by Gunn \& Griffin
(1979). We employed the multi-mass Michie--King code originally
developed by Meylan (1987, 1988) and later suitably modified by us
(Pulone, De Marchi \& Paresce 1999); De Marchi et al. 2000) for the
general case of clusters with multiple LF measurements at various
radii. Each model run is characterised by a MF in the form of a
power-law with a variable index $\alpha$, and by four structural
parameters describing, respectively, the scale radius ($ r_{\rm c}$),
the scale velocity ($v_{\rm s}$), the central value of the
dimensionless gravitational potential $W_{\rm o}$, and the anisotropy
radius ($r_{\rm a}$). 

From the parameter space defined in this way, we selected those models
that simultaneously fit both the observed surface brightness profile
(SBP) and the central value of the velocity dispersion as given,
respectively, by Trager et al. (1995) and Webbink (1985). However,
while forcing a good fit to these observables constrains the values of
$r_{\rm c}$, $v_{\rm s}$, $W_{\rm o}$, and $r_{\rm a}$, the MF can
still take on a variety of shapes. To break this degeneracy, we impose
the additional condition that the model LF agrees with that observed at
all available locations simultaneously. This, in turn, sets very
stringent constraints on the present GMF, i.e. on the MF of the cluster
as a whole. 

\begin{figure}
\centering
\resizebox{\hsize}{11.65cm}{\includegraphics{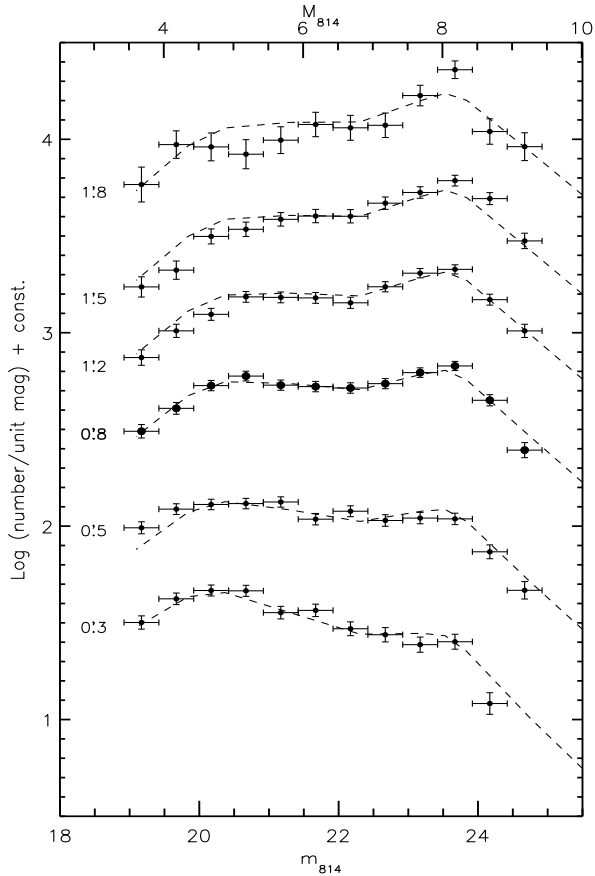}}
\caption{The LFs of Figure\,\ref{fig4} (circles) are compared with
those predicted by our multi-mass Michie--King model at various radii
inside the cluster (dashed lines), for a GMF with index $\alpha = 0.5$.
The expected radial variation of the LF is fully consistent with the
observations.}
\label{fig5}
\end{figure} 

For practical purposes, the model GMF has been divided into sixteen
different mass classes, covering main sequence stars, white dwarfs and
heavy remnants, precisely as described in Pulone et al. (1999). We ran
a large number of trials looking for a suitable shape of the GMF such
that the local MFs implied by mass segregation would locally fit the
observations. As explained in the previous section, in order to keep
observational errors and theoretical uncertainties separate, we
converted the model MFs to LFs, using the same M--L relation, distance
modulus and colour excess of Figure\,\ref{fig4}, and compared those to
the observations. 

Our analysis shows that the set of model LFs that best fits all
available observations simultaneously corresponds to a GMF index
$\alpha=0.5$ for stars less massive than $0.8$\,\Msolar. At higher
masses, we assume that the IMF had originally a power-law shape with
index $\alpha=-2$, close to the Salpeter value $-2.3$.  This parameter
determines the fraction of heavy remnants in the cluster (white dwarfs,
neutron stars and black holes), which we find to be of order $60\,\%$,
and affects the overall distribution of the stars in all other classes
of mass, to which the SBP is rather sensitive. We find that values of
$\alpha$ larger or smaller than $-2$ give a progressively worse
fit to the SBP, which becomes unacceptable for $\alpha > -1.5$ or
$\alpha < -2.5$.

\begin{figure}
\centering
\resizebox{\hsize}{!}{\includegraphics[width=16cm]{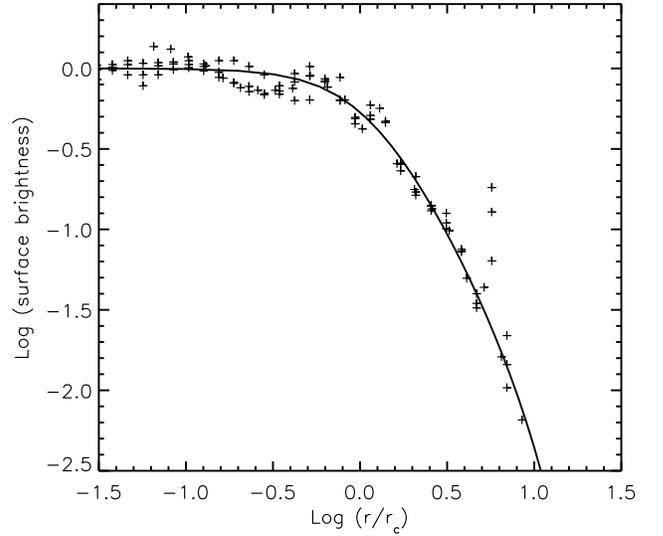}}
\caption{The surface brightness profile of NGC\,2298 (crosses, from
Trager et al. 1995) is well reproduced by our dynamical model (solid
line). The observations are normalised to the central value of the best
fitting profile as given by Trager et al. (1995).}
\label{fig6}
\end{figure}

The best fit to the LFs is shown in Figure\,\ref{fig5}. The same model
also reproduces remarkably well the SBP of Trager et al. (1995), as
shown in Figure\,\ref{fig6}, and the value of the core radius that we
obtain is $r_{\rm c}=18\arcsec$, in good agreement  with the literature
value $r_{\rm c}=20\arcsec$ of Harris (1996). We note here that in
Figure\,\ref{fig6} we compare to the data of Trager et al. (1995) the
model SBP for stars of $0.8$\,\Msolar, i.e. turn-off and red giant
stars, since these dominate the integrated light. Obviously, star of
different masses have a different radial distribution, with the
relative density at any location being governed by the relaxation
process (King 1966). The cluster's structural parameters and
integrated  luminosity derived from our best model are compared with
their corresponding literature values in Table\,\ref{tab4} and appear
in rather good agreement with the latter.

\begin{table}[b]
\caption[]{Cluster structural parameters for NGC\,2298}
\begin{tabular}{lccc}
\hline
Parameter & Fitted & Literature & Ref.\\ & value & value & \\
\hline
core radius        $r_{\rm c}$      & $0\farcm29$     & $0\farcm34$ & $a$\\
tidal radius       $r_{\rm t}$      & $8\farcm0$      & $6\farcm5$  & $a$\\
half-mass radius   $r_{\rm h}$      & $1\farcm2$      & $0\farcm8$  & $a$\\
central vel. disp. $\sigma_{\rm o}$ & $3.3$\,km/s     & $3.6$\,km/s & $b$\\
total luminosity   $M_{\rm V}$       & $-6.3$          & $-6.3$      &$a$\\
\hline
\end{tabular}
\par\noindent
$a$): Harris (1996)
\par\noindent
$b$): Webbink (1985)
\label{tab4}
\end{table}

By design, also the central velocity dispersion predicted by the model
matches the literature value (see Table\,\ref{tab4}), although the
latter is not well constrained for NGC\,2298. The value published by
Webbink (1985), $\sigma_{\rm o}=3.6$\,km/s, is not a direct
measurement, but rather the result of a model. To our knowledge, the
only recent radial velocity measurements in NGC\,2298 are those by
Geisler et al. (1995) of 9 red giant stars located between $1^\prime$
and $1\farcm5$ from the cluster centre, or approximately at the
half-mass radius. These measurements have an accuracy of $\sim 2$\,km/s
and their standard deviation is $3.7$\,km/s. This value is in agreement
with the estimate of Webbink (1985). It also agrees with the prediction
of our model, which gives a velocity dispersion of $\sim 3$\,km/s
between $1^\prime$ and $1\farcm5$ from the centre. 

From $r_{\rm c}$, $\sigma_{\rm o}$ and $W_{\rm o}$ we derive the total
cluster mass to be $5.1 \times 10^4$\,\Msolar.

The good agreement between the observed radial variation of the LF and
that predicted by the Michie--King model (Figure\,\ref{fig5}) implies
that stars in NGC\,2298 must be very close to a condition of
equipartition, at least over the area covered by the observations.
Sosin (1997) has shown that this is the case even if the phase space
distribution function may not be the lowered-Maxwellian required by the
Michie--King model (Michie \& Bodenheimer 1963; King 1966). The
half-mass relaxation time that our model indicates is $t_{\rm
hm}=3.1$\,Gyr, considerably shorter than GC ages (Krauss \& Chaboyer
2003). As for the shape of the GMF, the index $\alpha=0.5$ implies that
it is decreasing with mass. This confirms that the deficiency of
low-mass stars inferred from Figure\,\ref{fig4} is a general
characteristic of NGC\,2298 and not just the local effect of mass
segregation. The excellent match between the GMF index $\alpha=0.5$ and
that of the local MF measured near the half-mass radius (also
$\alpha=0.5$) shows once more that the latter is an accurate
approximation to the cluster's GMF (Richer et al. 1991; De Marchi et
al. 1995; De Marchi et al. 2000).

\section{Discussion and conclusions}

According to the classification of Zinn (1993), NGC\,2298 is an old
halo cluster by virtue of its low metallicity [Fe/H]$=-1.85$ (Harris
1996). The age of $\sim 12$\,Gyr estimated by Testa et al. (2001) from
the properties of horizontal branch stars is fully consistent with this
classification. As indicated above, NGC\,6397, another old halo
cluster, has very similar metallicity [Fe/H]$=-1.95$. Its age is also
similar: Gratton et al. (2003), using isochrone fitting to the MS,
estimate $13.4 \pm 0.8 \pm 0.6$\,Gyr and Richer et al. (2005), using
the white dwarf cooling sequence, derive $\sim 12.7$\,Gyr, in good
agreement. It appears, therefore, logical to suppose that the two
clusters were born with a similar IMF. 

Their present GMFs, however, are quite different. As noted in
Section\,5, that of NGC\,2298 is best described by a power-law with
index $\alpha=0.5$ over the range $0.2-0.8$\,\Msolar. As for NGC\,6397,
although its GMF is more complex and is best represented by a
log-normal distribution (PDM00) or by a tapered power-law (De Marchi et
al. 2005), over the mass range $0.3 - 0.8$\,\Msolar\, it is still well
described by a simple power-law of index $\alpha=-1.6$ (De Marchi et
al. 2000).\footnote{Note that, since NGC\,2298 and NGC\,6397 have a
very similar metallicity, the two GMFs were determined by using the
same M--L relationship. Therefore, any differences in the resulting
GMFs reflect the physical properties of the clusters and are rather
insensitive to the uncertainties in the theoretical M--L models.} Thus,
the marked difference in their present GMF (PGMF) must be the result of
dynamical evolution, namely the evaporation of stars via two-body
relaxation, possibly enhanced by the effect of the Galactic tidal field
(Aguilar, Hut \& Ostriker 1988; Vesperini \& Heggie 1997; Murali \&
Weinberg 1997; Gnedin \& Ostriker 1997).

Since we have studied the present dynamical structure of both clusters
(see Section\,5 above and De Marchi et al. 2000), we can look for signs
of a faster mass loss from NGC\,2298. Both objects are in a condition
of equipartition, as shown by the fact that the radial variation of
their MFs resulting from mass segregation is well fitted by a simple
multi-mass Michie--King model. They also have a similar total mass,
namely $5.1 \times 10^4$\,\Msolar\, for NGC\,2298 and $8.9 \times
10^4$\,\Msolar\, for NGC\,6397 (De Marchi et al. 2000). We note here
that both mass values come from a multi-mass Michie--King model and are
based on measured velocity dispersions rather than by simple
multiplication of the total cluster luminosity by an arbitrary $M/L$
ratio, as one too often finds in the literature. Their density,
however, is rather different, as witnessed by the value of the central
concentration parameter $c=\log(r_{\rm t}/r_{\rm c})$, which takes on
the value of $1.3$ and $2.5$, respectively for NGC\,2298 and NGC\,6397
(Harris 1996). 

In fact, the most outstanding dynamical difference between the two
objects is that NGC\,6397 appears to be in a more advanced evolutionary
stage, since its high central concentration suggests that  it has a
post-collapse core (King, Sosin \& Cool 1995). This, however, should
not imply a higher mass-loss rate for NGC\,2298. In fact, one would
expect clusters in a more advanced dynamical phase (i.e. those with
post-collapse cores) to have lost more low-mass stars via the
evaporation process caused by two-body relaxation (Vesperini \& Heggie
1997).

In any case, the evaporation process alone appears to be insufficient
to account for the depletion of low mass stars in NGC\,2298 and, as
such, it may not be the primary cause of mass loss from this cluster.
We show this with a simple ``order-of-magnitude'' calculation. If we
assume that NGC\,2298 and NGC\,6397 were born with the same IMF and,
following the conclusions of PDM00, that the PGMF of NGC\,6397 is
substantially coincident with its IMF, then a simple integration of the
difference between the two PGMFs shows that NGC\,2298 must have lost
about 85\,\% of its original mass in the course of its life. This would
imply an average evaporation rate of $\sim 2 \times
10^{-5}$\,\Msolar/yr. 

Evaporation takes place on a timescale considerably longer than
relaxation, at least one order of magnitude longer (Spitzer 1987).
Based on the value of the half-mass relaxation time that we find
($t_{\rm rh}\simeq 3$\,Gyr), we estimate an evaporation timescale
$t_{\rm evap}\ga 30$\,Gyr, for the present mass and structure of the
cluster, corresponding to an average evaporation rate of $\la 2 \times
10^{-6}$\,\Msolar/yr, or an order of magnitude less than that required
to deplete the GMF of NGC\,2298 at the low mass end to the level
observed today. Considering that $t_{\rm evap}$ scales with the total
mass of the cluster, and was therefore longer in the past, it appears
unlikely that the internal evaporation process alone may have removed a
conspicuous fraction of NGC\,2298's original mass in the $\sim 13$\,Gyr
of its life.

Clearly, however, evaporation is not the only process that can remove
stars from a GC. Clusters with orbits crossing the Galactic plane
or venturing close to the Galactic centre undergo compressive heating,
respectively via disc and bulge shocking, which can have a much
stronger effect than evaporation on the loss of stars, depending
on the orbit (Aguilar, Hut \& Ostriker 1988; Gnedin \& Ostriker 1997;
Dinescu et al. 1999b). 

Orbital parameters have been determined for both clusters via proper
motion and radial velocity measurements (Dinescu et al. 1999a, 1999b;
Cudworth \& Hanson 1993; Dauphole et al. 1996). Here a remarkable
difference exists between NGC\,6397, which occupies a low-eccentricity
orbit ranging between 3 and 6\,kpc from the Galactic centre, and
NGC\,2298, which has a very eccentric orbit with perigalactic and
apogalactic distances of respectively 2 and 15\,kpc. The maximum
distance from the Galactic plane for NGC\,6397 is $1.5$\,kpc, whereas
NGC\,2298 reaches as far as $\sim 7$\,kpc. The orbital periods are 143
and 304\,Myr, respectively (Dinescu et al. 1999b). It appears that, for
most of its orbit, NGC\,2298 is too far removed from the disc and bulge
of the Galaxy for the compressive heating mentioned above to be a
serious threat for the cluster. Indeed, on the basis of these space
motion parameters, Gnedin \& Ostriker (1997), Dinescu et al. (1999b)
and Baumgardt \& Makino (2003) conclude that evaporation is the
dominant cause of disruption for both clusters. Quite surprisingly, the
rate of disruption of NGC\,6397 is twice as high as that of NGC\,2298,
according to these authors. This is not compatible with the
observations, if the clusters were born with a similar IMF.

The careful reader will not have missed that it is also entirely
possible that our assumption is ill founded and that NGC\,2298 and
NGC\,6397 were not born with the same IMF, even though they have
similar age and metallicity. In fact, given the evidence above and
taking the model predictions at face value, one would be tempted to
conclude that NGC\,2298 was born with a considerably flatter IMF (i.e.
with a smaller proportion of low-mass stars) than NGC\,6397 or any
other known halo cluster (PDM00). As explained in De Marchi, Paresce \&
Pulone (2007), however, the existence of a trend between the central
concentration of a cluster and its GMF index argues against this
hypothesis. Denser clusters are found to have a systematically steeper
GMF than loose clusters and since it is hard to imagine that the star 
formation process could know about the final dynamical structure of 
the forming cluster, the observed trend is most likely the result of 
dynamical evolution, i.e. low-mass star evaporation and tidal
stripping. Therefore, the origin of the large difference between the
GMFs of NGC\,6397 and NGC\,2298 will have to be searched in the details
of the past dynamical histories of these two objects, which are still
largely unknown.

To be sure, models of how GCs interact with the Galaxy are rather
uncertain and their predictions have to be taken with great care.
Uncertainties on the orbit of individual clusters, determined from
their current proper motion and radial velocity, can be large.
Furthermore, the actual distribution of the mass in the Galaxy has
never been directly  measured and models must be used instead (e.g. 
Bahcall, Soneira \& Schmidt 1983; Ostriker \& Caldwell 1985; Johnston,
Spergel \& Hernquist 1995), which are affected by unavoidable
uncertainties. Therefore, while the predictions of how GCs interact
with the Galaxy may well be valid in a statistical sense when it comes
to describing the global properties of the GC system, their
applicability to individual objects is not guaranteed (De Marchi et al.
2006). For instance, the perigalactic distance of NGC\,2298 is known
with considerable uncertainty, $R_{\rm p} = 2.1 \pm 1.6$\,kpc, compared
to NGC\,6397 for which $R_{\rm p} = 3.1 \pm 0.2$\,kpc (Dinescu et al.
1999a). Since the rate of disruption due to bulge shocking scales with
$R_{\rm p}^{-4}$, if $R_{\rm p}$ for NGC\,2298 were to decrease by a
factor of 2 or 3 (compatible with the present uncertainty), bulge
shocking could become the dominant mass loss mechanism and explain, at
least qualitatively, the difference with the PGMF of NGC\,6397.
Unfortunately, this situation will not improve until more accurate
astrometric information becomes available for a sizeable number of GCs
and a better understanding of the mass distribution in the bulge is
secured. It is expected that this will be possible with the advent of
Gaia.

\begin{acknowledgements}

We thank an anonymous referee for very constructive comments that have
helped us to improve the presentation of our work. GDM is grateful to
ESO for their hospitality via the Science Visitor Programme during the
preparation of this paper. The work of LP was partially supported by
programme PRIN-INAF 2005 (PI: M. Bellazzini), ``A hierarchical merging
tale told by stars: motions, ages and chemical compositions within
structures and substructures of the Milky Way''.

\end{acknowledgements}

\end{document}